# Multistability in Networks of Weakly Coupled Bistable Units

R.S. MacKay [a] and J.-A. Sepulchre [a,b]

[a] *Nonlinear Systems Laboratory, Mathematics Institute, University of Warwick, CV4 7AL Coventry, United Kingdom*

[b] *previous address: Université Libre de Bruxelles, CP 231, Boulevard du Triomphe, B-1050 Bruxelles, Belgium*

**Abstract**

We study the stationary states of networks consisting of weakly coupled bistable units. We prove the existence of a high multiplicity of stable steady states in networks with very general inter-unit dynamics. We present a method for estimating the critical coupling strength below which these stationary states persist in the network. In some cases, the presence of time-independent localized states in the system can be regarded as a 'propagation failure' phenomenon. We analyse this type of behaviour in the case of diffusive networks whose elements are described by one or two variables and give concrete examples.

## 1  Introduction

One of the simplest behaviours in nature which reflects underlying nonlinear mechanisms, is the phenomenon of *bistability*, that is the possibility for a system to present simultaneously two stationary states which are stable for the same set of system parameters. This type of behaviour has been identified in a great number of natural systems, ranging from laser physics to cellular biology.

In physics, examples of bistable phenomena are the behaviour of a metallic strip under longitudinal squeeze, the input-output response of diode circuits [1], the onset of electrical filaments in gas-discharge experiments [2], and the phase separation dynamics in condensed matter [3]. Another interesting example of bistability is studied in nonlinear optics [4]. For example, if a coherent electrical field is injected into a ring cavity containing an absorbing medium, the latter can respond in two different ways for the same amplitude of the driving field. In the first regime, the energy which is transmitted into the cavity is nearly totally reflected, whereas the second regime is characterized by a high transmission rate. Such devices have been proposed for building optical memories.

Other phenomena of bistability have been demonstrated in chemical systems. For instance, the bistability of the chloride-iodite reaction has been studied over a wide range of external constraints [5]. In biology also, interesting cases of bistability have been identified [6], with, for example, the membrane potential of neural cells which can sustain, under some circumstances, two stable equilibria. This latter situation can be described simply by the FitzHugh-Nagumo equations [6].

In most of these systems, the mathematical description of the bistability phenomena is fairly well understood, and it involves typically a nonlinearity of the 'cubic-type'. What is still under investigation, however, is the understanding of the possible behaviour that bistable elements exhibit, when coupled together in a large network. In this paper, we concentrate on the behaviour which appears in a network of coupled bistable units, for small value of the coupling strength between units. Using the implicit function theorem we show that, for a wide class of interaction between network units, the individual bistable behaviour of the units persists in the system, if these units are weakly coupled. It gives rise to a high *multistability* of the network stationary states. Some of these are very localized, for example, the continuations of the states for which all the units are in the same equilibrium, except for one unit which is in the other equilibrium.

Many of the steady states of the network may disappear as the coupling strength is increased. Numerical experiments typically show that in this case, one can observe the onset of an activity front which propagates in the network. On the other hand, when the coupling strength is below a critical value, propagation becomes impossible, and we suggest that the presence of steady localized structures can be interpreted as *propagation failure*. This situation has been studied recently by several authors in the case where the interaction between units of the network is described by discrete diffusion [8]-[12]. For example, Keener [8] studied the propagation failure which can occur in a diffusive chain of excitable cells. He proposed a method for obtaining an analytical estimate of the critical coupling strength below which propagation is blocked. This method is valid for one-dimensional translation invariant discrete reaction-diffusion systems with one variable per unit. In the present paper, we propose a technique which allows one to extend this kind of estimation to networks with more general connectivity, not necessarily translation invariant and whose units are described by more than one variable.

The paper is organized as follows. In section 2, we formulate the problem as one of continuation of steady states from the uncoupled case. Section 3 proposes a general scheme for estimating a critical coupling strength up to which all the steady states persist. Section 5 applies this scheme to networks described by two variables per unit interacting by means of discrete diffusion. In section 6, the same technique is applied for studying the persistence of steady states in networks consisting of bistable optical devices. In section 7 we discuss the stability-type of the resulting steady states. Finally, section 8 is devoted to conclusions.

## 2 Continuation from the uncoupled limit

In this section we show how steady states of an uncoupled network of bistable units can be continued to weak coupling, for a range of coupling which depends only on bounds on the dynamics of the individual units and the $C^1$ norm of the coupling, regarded as an operator on a sup-norm Banach space. We emphasize that the size of the network does not enter explicitly.

We consider a network consisting of a large and possibly infinite number of units coupled together. The units of the network are indexed by an index set $\mathbf{S}$ (e.g. the integers $\mathbf{Z}$ or $\mathbf{Z}^2$), and the state of the unit $n \in \mathbf{S}$ is described by a vector $x_n = (v_n, \ldots, w_n) \in \mathbf{R}^m$. Then, a state of the whole network is defined by an indexed set $X = (x_n)_{n \in \mathbf{S}}$ of $m$-dimensional vectors. We choose a norm $|\cdot|$ on $\mathbf{R}^m$, and extend it to the set of bounded states of the network by

$$|X| = \sup_n |x_n| \, .$$

With this norm, the set of bounded states of the network forms a Banach space that we denote by $\mathbf{R}^{m\mathbf{S}}$. Now we define the dynamics of the network.

First we assume that the dynamics of each unit taken separately is described by the differential equation

$$\frac{\mathrm{d}x_n}{\mathrm{d}t} = f(x_n) \, , \tag{2.1}$$

where the $C^1$ function $f : \mathbf{R}^m \to \mathbf{R}^m$ has the property of bistability. This means that the equation $f(x_n) = 0$ has two distinct solutions $x_n = x^{(i)}$ ($i = 0, 1$) such that all eigenvalues of the matrix $\mathrm{D}f(x^{(i)})$ have negative real part.

Next we define the map $F : \mathbf{R}^{m\mathbf{S}} \to \mathbf{R}^{m\mathbf{S}}$ by

$$F(X) = (f(x_n))_{n \in \mathbf{S}} \, .$$

Then we assume that the dynamics of the network of bistable units obeys a differential equation of the type

$$\frac{dX}{dt} = F(X) + \alpha\, K(X)\,, \tag{2.2}$$

where $\alpha$ is a real parameter called the coupling strength and $K : \mathbf{R}^{m\mathbf{S}} \to \mathbf{R}^{m\mathbf{S}}$ is a $C^1$ function which describes the interaction between units. This interaction function is possibly nonlinear. An example for the function $K$, of special interest in physics, is the discrete diffusion operator in a $d$-dimensional cubic lattice. In that case, each component of $K$ is a linear function

$$K_n(X) = D \sum_{k=1}^{2d} (x_{n_k} - x_n)\,, \tag{2.3}$$

where $D$ is a $m \times m$ diagonal matrix specifying the 'diffusion coefficients', and the subscripts $k$ in eq. (2.3) run over the $2d$ nearest neighbours of the unit $x_n$ in the lattice. For $d = 1$, it gives the well-known finite difference $x_{n+1} - 2x_n + x_{n-1}$.

Our analysis starts from the uncoupled case, $\alpha = 0$. This provides explicitly a large number of stationary states. Indeed, for any choice of binary indexed set $(i_k)_{k\in\mathbf{S}}$, $i_k = 0$ or $1$, there exists a corresponding stationary state of eq. (2.2) given by

$$X_0 = (x^{(i_k)})_{k\in\mathbf{S}}\,. \tag{2.4}$$

It satisfies the equation $F(X_0) = 0$.

Our idea is that all of these steady states can be continued to steady states for small $\alpha$. The strategy is borrowed from [7].

**Theorem 2.1** *Given a system of the form (2.2), $\exists\ \alpha_0 > 0$ such that any stationary state $X_0$ of the uncoupled case has a locally unique continuation for all $|\alpha| < \alpha_0$.*

**Proof.** This theorem is proved by a simple application of the implicit function theorem. Let us rewrite the right-hand side of eq. (2.2) as a map $G : \mathbf{R}^{m\mathbf{S}} \times \mathbf{R} \to \mathbf{R}^{m\mathbf{S}}$ defined by

$$G(X, \alpha) = F(X) + \alpha\, K(X)\,. \tag{2.5}$$

The stationary state $X_0$ is such that $G(X_0, 0) = 0$. The zero $(X_0, 0)$ of $G$ has a locally unique continuation around $\alpha = 0$ if, by the implicit function theorem, the derivative of $G$ with respect to $X$, denoted by $DG$, is invertible at $(X_0, 0)$ (with bounded inverse) and the map $G$ is continuously differentiable.

Now, $DG(X_0, 0) = DF(X_0)$ is block diagonal with blocks $J_{i_k} = Df(x^{(i_k)})$, and each block is invertible since all the eigenvalues of $J_{i_k}$ have strictly negative real part. Hence $DG^{-1}(X_0, 0)$ exists and is block-diagonal, with blocks equal to $J_{i_k}^{-1}$. Since the norms of the blocks are bounded, $DG^{-1}$ is bounded. $F$ is $C^1$, since $f$ is $C^1$. $K$ is $C^1$ by assumption. Hence $G$ is $C^1$. Therefore there exists an interval of values of $\alpha$ around 0 such that the equation $G(X, \alpha) = 0$ defines a function $X(\alpha)$.

A common interval $[-\alpha_0, \alpha_0]$ can be found which works for all choices of steady state $X_0$ because the estimates in the implicit function theorem can be chosen independently of the choice of $X_0$. □

**Remark 2.2** The function $X(\alpha)$ is differentiable with respect to $\alpha$ and

$$\frac{dX}{d\alpha} = -DG^{-1}(X, \alpha) K(X) , \qquad (2.6)$$

as long as $DG(X, \alpha)$ is invertible. In the next section we will estimate a lower bound on $\alpha$ for which $DG^{-1}$ exists.

**Remark 2.3** The result of theorem 2.1 is independent of the specific form of the interaction function $K(X)$ (provided it is $C^1$). Consequently, the theorem is valid for a class of network interactions which is much wider than the discrete diffusion defined by eq. (2.3). Note that when $K$ is linear, the $C^1$ condition is just that the sum of the norms of the elements along the rows of the matrix $K$ be bounded.

**Remark 2.4** Theorem 2.1 remains true even if the local dynamics $f$ varies a little from unit to unit. All that is required is uniform bounds on $Df$ and $(Df)^{-1}$ in a neighbourhood of the equilibria of $f$.

## 3   Procedure for estimation of $\alpha_0$

Theorem 2.1 means that in a network of weakly coupled bistable units, there exist a great number of nonuniform states of the system which do not evolve in time. Stability is proved in section 7.

The persistence of all the stationary states will occur as long as the operator $DG(X, \alpha)$ remains invertible. The goal of this section is to estimate a value $\alpha_0$ of the coupling parameter up to which all the stable steady states persist. It is possible, and indeed likely, that they will persist further than $\alpha_0$. Furthermore, the true threshold in $\alpha$ for continuation of a steady state will in general vary from steady state to steady state. So $\alpha_0$ should be interpreted just as a lower bound on the set of critical $\alpha$'s for all the steady states.

The method is based on the estimation of the norms of each sides of eq. (2.6). Let us define the variable

$$\zeta = |X - X_0| , \tag{3.1}$$

If $\|\mathrm{D}G^{-1}(X,\alpha)\|$ can be bounded by a function $P(\zeta,\alpha)$ and $\|K(X)\|$ by a function $Q(\zeta)$, then the solution $X_0$ can be continued with

$$\left|\frac{\mathrm{d}\zeta}{\mathrm{d}\alpha}\right| \leq P(\zeta,\alpha)Q(\zeta) ,$$

as long as $P(\zeta,\alpha) < \infty$. Let $\alpha_0$ be the first $\alpha$ such that $P$ becomes infinite along the solution of

$$\frac{\mathrm{d}\zeta}{\mathrm{d}\alpha} = P(\zeta,\alpha)Q(\zeta) , \tag{3.2}$$

starting from $\zeta(0) = 0$. Then the solution $X(\alpha)$ can be continued at least up to $\alpha_0$.

To implement this procedure, we first note a useful bound for $\|\mathrm{D}G^{-1}\|$. By definition of the function $G$ we have

$$\mathrm{D}G = \mathrm{D}F + \alpha\,\mathrm{D}K . \tag{3.3}$$

So if $\mathrm{D}F$ is invertible and $|\alpha|\,\|\mathrm{D}K\| < \|\mathrm{D}F^{-1}\|^{-1}$ then $\mathrm{D}G$ is invertible and

$$\|\mathrm{D}G^{-1}\| \leq \frac{1}{\|\mathrm{D}F^{-1}\|^{-1} - |\alpha|\,\|\mathrm{D}K\|} . \tag{3.4}$$

This is easy to use because $\mathrm{D}F$ is block diagonal, so $\|\mathrm{D}F^{-1}\|^{-1}$ is the minimum of the quantities $\|\mathrm{D}f(x_n)^{-1}\|^{-1}$. Sometimes it is useful to obtain a slightly stronger bound on $\mathrm{D}G^{-1}$ by incorporating the diagonal part of $\mathrm{D}K$ into $\mathrm{D}F$, i.e., write

$$\mathrm{D}K = L_0 + L_1 , \tag{3.5}$$

where $L_0$ is the block diagonal (i.e. on-site) part of the operator $\mathrm{D}K$ and $L_1 = \mathrm{D}K - L_0$ is off-diagonal (i.e. inter-site part). Then the operator $\mathrm{D}G$ is decomposed as

$$\mathrm{D}G = \Lambda(\alpha) + \alpha L_1 , \tag{3.6}$$

where $\Lambda(\alpha) = \mathrm{D}F + \alpha L_0$ is block-diagonal by construction. Thus the inverse $\Lambda^{-1}$ is also block-diagonal, with blocks equal to $(\mathrm{D}f(x_n) + \alpha(L_0)_n)^{-1}$ that we assume to be bounded because this is true for small $\alpha$ by our hypothesis on the spectrum of $\mathrm{D}f(x^{(i)})$. Then, provided $|\alpha| \parallel L_1 \parallel < \parallel \Lambda^{-1} \parallel^{-1}$, the norm of $\mathrm{D}G^{-1}$ can be bounded by

$$\parallel \mathrm{D}G^{-1} \parallel \leq \frac{1}{\parallel \Lambda^{-1}(\alpha) \parallel^{-1} - |\alpha| \parallel L_1 \parallel} . \tag{3.7}$$

After this, there are many ways of proceeding the bounds $\parallel \Lambda^{-1}(\alpha) \parallel^{-1}$, $\parallel L_1 \parallel$ and $|K|$. If the system has special features, one can obtain nice bounds on these quantities and then solve eq. (3.2) explicitly. We will give an example in section 4. If not, one can nevertheless always obtain bounds of the following form, for $\zeta \leq$ some $\zeta_1$, since $F$ and $K$ are $C^1$:

$$\parallel \Lambda^{-1}(\alpha) \parallel^{-1} \geq a_0 - a_1 \zeta - b_1 |\alpha| , \tag{3.8}$$
$$\parallel L_1(X) \parallel \leq b_0 , \tag{3.9}$$
$$|K(X)| \leq c_0 + c_1 \zeta , \tag{3.10}$$

with $a_0, a_1, b_0, b_1, c_0$ and $c_1$ real numbers, $a_0, b_0, c_0 > 0$. Therefore, considering the norm of each side of eq. (2.6), and combining eq. (3.7)-(3.10), we obtain the differential inequality

$$\left| \frac{\mathrm{d}\zeta}{\mathrm{d}\alpha} \right| \leq \frac{c_0 + c_1 \zeta}{a_0 - a_1 \zeta - b|\alpha|} , \tag{3.11}$$

with $b = b_0 + b_1$. Starting with the initial condition $\zeta(0) = 0$, eq. (3.11) holds as long as the denominator remains positive and $\zeta < \zeta_1$. The case of equality in eq. (3.11) can be integrated by introducing a new variable $s$ in order to separate the numerator and denominator of the fraction (i.e., replace $\frac{\mathrm{d}\zeta}{\mathrm{d}\alpha}$ by $\frac{\mathrm{d}\zeta}{\mathrm{d}s}/\frac{\mathrm{d}\alpha}{\mathrm{d}s}$). The integration of the resulting equation permits one to determine the point $(\zeta_0, \alpha_0)$ for which the denominator of eq. (3.11) vanishes for the first time. If we assume $\zeta_0 \leq \zeta_1$, we obtain

$$\alpha_0 = \frac{a_0}{b} + \frac{a_1 c_0}{b c_1} \left( 1 - [1 + \frac{a_0(c_1 + b)}{a_1 c_0}]^{c_1/(c_1+b)} \right)$$

It gives a lower bound on $\alpha$ up to which $\mathrm{D}G(X(\alpha), \alpha)$ remains invertible. Using the binomial expansion, it is instructive to expand $\alpha_0$ as the series:

$$\alpha_0 = \frac{1}{2} \frac{a_0^2}{a_1 c_0} - \frac{(2b + c_1)}{6} \frac{a_0^3}{a_1^2 c_0^2} + \mathcal{O}(\frac{a_0^4}{a_1^3 c_1^3}) \tag{3.12}$$

We note that the parameters $b$ and $c_1$ enter only from the second term.

To end this section we notice that the solution of eq. (2.6) is not affected by the multiplication of $\mathrm{D}G$ and $K$ by a common regular matrix $M$. Nonetheless the relation (3.11), which is in fact derived from eq. (2.6), can be altered by such a matrix multiplication. This property can be used in order to optimize the estimation represented by the inequality (3.11). In the next section we apply the scheme presented above in the case of bistable elements described by one variable and when the interaction between units is described by a diffusion-like operator, as defined by eq. (2.3).

## 4  One variable systems

To demonstrate the method, we begin by treating the simplest case: one-dimensional translation-invariant nearest-neighbour-coupled one-variable systems:

$$\dot{v}_n = f(v_n) + \alpha \left(v_{n+1} + v_{n-1} - 2v_n\right), \tag{4.1}$$

with $v_n \in \mathbf{R}$ and $n \in \mathbf{Z}$. We restrict attention to $\alpha \geq 0$ as this is the case of most physical interest. We suppose $f$ has (at least) two stable equilibria, which can be taken without lost of generality at $v^{(0)} = 0$ and $v^{(1)} = 1$ (with $f'(0), f'(1) < 0$). Let $I_0, I_1$ be the largest intervals about 0, respectively 1, on which $f'(v) < 0$. We illustrate (3.7), and the special form of this problem will allow us to find good estimates for (3.8)-(3.10).

The blocks of $\Lambda$ are $f'(v_n) - 2\alpha$. So, if $v_n \in I_0 \cup I_1$, for all $n$, and if $\alpha \geq 0$ then

$$\| \Lambda^{-1} \|^{-1} = \inf_{v_n} |f'(v_n)| + 2\alpha$$

Now $\| L_1 \| = 2$. Thus eq. (3.7) gives

$$\| \mathrm{D}G^{-1}(v, \alpha) \| \leq \frac{1}{\inf_{v_n} |f'(v_n)|}, \qquad \text{for } v_n \in I_0 \cup I_1, \tag{4.2}$$

and for $\alpha \geq 0$. Note the cancellation of the "$2\alpha$"s which is convenient, though not essential. Next, in the present case $K_n(v) = (v_{n+1} + v_{n-1} - 2v_n)$ and

$$|K(v)| \leq 2 + 4\zeta ,\qquad$$

with $\zeta$ defined by (3.1). We can do slightly better than this, however, because all equilibrium states for $\alpha \geq 0$ satisfy

$$f(\underline{v}) \leq 0 , \qquad f(\overline{v}) \geq 0 , \tag{4.3}$$

where $\underline{v} = \inf v_n$ and $\overline{v} = \sup v_n$. To prove this, note that the equation for an equilibrium state is

$$f(v_n) = -\alpha \left(v_{n+1} + v_{n-1} - 2v_n\right) .$$

Now, the relations (4.3) and the negative slope of $f$ at $v = 0$ and $1$ imply that the equilibrium states obtained by continuation from $v_n \in \{0, 1\}$ (for all $n$) satisfy

$$v_n \in [0, 1] \qquad \text{for all } n .$$

Hence along continuation we have

$$|K(v)| \leq 2 . \tag{4.4}$$

Combining eqs. (4.2)-(4.4) we deduce that the continuation can be performed with

$$\frac{\mathrm{d}\zeta}{\mathrm{d}\alpha} \leq \frac{2}{D(\zeta)} , \tag{4.5}$$

with the function $D(\zeta)$ defined as

$$D(\zeta) = \inf_{v \in [0,\zeta] \cup [1-\zeta,1]} |f'(v)| ,$$

as long as the denominator $D(\zeta)$ remains positive. The inequality (4.5) is easy to solve and thus continuation is possible at least as long as $\alpha < \alpha_0$ with

$$\alpha_0 = \frac{1}{2} \int_0^{\zeta_0} D(\zeta) \, \mathrm{d}\zeta , \tag{4.6}$$

where $\zeta_0$ is the first value of $\zeta$ such that $D(\zeta) = 0$. Equivalently, $\zeta_0$ is the distance from $\{0, 1\}$ to the nearest critical point $c$ of $f$.

This result holds whether $f'(v)$ is monotonous or not. If $f'(v)$ is monotonous on $[0, \zeta_0]$ and on $[1-\zeta_0, 1]$, then $D(\zeta) = \min(|f'(\zeta)|, |f'(1-\zeta)|)$. Different cases may be distinguished. The simplest case is when $c < \frac{1}{2}$ and $|f'(\zeta)| \leq |f'(1-\zeta)|$ for all $\zeta \in [0, c]$. Then

$$\int_0^{\zeta_0} D(\zeta)\, d\zeta = \int_0^c -f'(\zeta)\, d\zeta = -f(c)\,.$$

A similar case with $c > 1/2$ is treated analogously. Thus, in this simplest situation, continuation is possible for

$$\alpha < \alpha_0 = \begin{cases} -\frac{f(c)}{2} & \text{if } c < \frac{1}{2} \\ \frac{f(c)}{2} & \text{if } c > \frac{1}{2} \end{cases} \tag{4.7}$$

In other words, our estimate of $\alpha_0$ is the half height of the nearest critical point of $f$ to 0 or 1.

Another example is to suppose $c < \frac{1}{2}$ but

$$|f'(\zeta)| \geq |f'(1-\zeta)| \text{ for } \zeta \in [0, \zeta_m]\,, \tag{4.8}$$
$$|f'(\zeta)| \leq |f'(1-\zeta)| \text{ for } \zeta \in [\zeta_m, c]\,.$$

Then, integral of $D(\zeta)$ can be calculated again and (4.6) gives

$$\alpha_0 = \frac{1}{2}\left(-f(c) + f(\zeta_m) + f(1-\zeta_m)\right)\,. \tag{4.9}$$

We will see an example of that type in section 6.

**Example 4.1** A typical example of a simple model of one-variable bistable dynamics is the cubic Nagumo model [6]. In this case $f$ can be defined as a cubic polynomial $p(v)$ of the form

$$p(v) = v(v-\eta)(1-v), \quad \text{with } 0 < \eta < 1\,. \tag{4.10}$$

The equilibria 0 and 1 of $p$ are stable as $p'(0) = -\eta < 0$ and $p'(1) = 1 - \eta < 0$. We restrict attention to $0 < \eta < \frac{1}{2}$ (the complementary case is treated similarly). Then the nearest critical point of $p$ to $\{0, 1\}$ is

$$c(\eta) = \frac{1}{3}\left(\eta + 1 - \sqrt{1 + \eta(\eta - 1)}\right) \simeq \frac{\eta}{2}.$$

Furthermore it can be shown that in the present case $|p'(\zeta)| < |p'(1 - \zeta)|$ for $\zeta < \frac{1}{2}$. Hence, the result expressed by eq. (4.7) applies and the continuation from the uncoupled limit of cubic Nagumo units is possible at least for $\alpha < \alpha_0(\eta) = -\frac{p(c(\eta))}{2}$. The analytical expression of $\alpha_0(\eta)$ is rather complicated but we can expand it as a power series in $\eta$:

$$\alpha_0 = \frac{\eta^2}{8} - \frac{\eta^3}{16} + \mathcal{O}(\eta^4). \tag{4.11}$$

This value of $\alpha_0$ is in agreement, at the lowest order, with the value $\eta^2/8$ predicted by Keener with a different approach [8]. The method proposed by Keener relies on a Smale's horseshoe type construction. Although result (4.11) is not quite as good as Keener's [8], our method generalises easily to many systems inacessible to Keener's method. For example, take $\mathbf{S}$ to be a translation invariant lattice where each site has $N$ neighbours with discrete diffusion

$$K_n(v) = \sum_{j=1}^{N}(v_{n_j} - v_n)$$

where $n_j$ are the neighbouring sites to $n$ (e.g. $N = 2d$ for a $d$-dimensional cubic lattice). Then the only change required to the above analysis is to replace the "2" in eqs. (4.4)-(4.6) by "N".

Extensions to one-variable systems on non-translation invariant lattices or with other forms of coupling are also relatively straightforward. For example, take $\mathbf{S}$ to be any graph, and

$$K_n(v) = \sum_j D_{nj}(v_j - v_n),$$

with $\{D_{nj} > 0; n, j \in \mathbf{S}\}$ a set of real numbers characterising the coupling between the bistable units of the system. As a generalisation of eq. (4.1), we consider the network of coupled bistable units described by the system:

$$\dot{v}_n = f(v_n) + \alpha K_n(v),$$

with $\alpha > 0$. Then, doing the same reasoning as for eqs. (4.2)-(4.5), we obtain the following differential inequality:

$$\left|\frac{\mathrm{d}\zeta}{\mathrm{d}\alpha}\right| \leq \frac{M}{D(\zeta) - (M-m)\alpha} \,, \tag{4.12}$$

where $M = \max_n \sum_j D_{nj}$ and $m = \min_n \sum_j D_{nj}$. The case of equality is easily integrated and we can express $\alpha_0$ under the form

$$\alpha_0 = \frac{1}{M} \int_0^{\zeta_0} \mathrm{e}^{-(1-\frac{m}{M})(\zeta_0 - \zeta)} D(\zeta) \,\mathrm{d}\zeta \,,$$

where $\zeta_0$ is the first value of $\zeta$ for which the denominator of (4.12) vanishes. Extensions to networks where the local dynamics $\dot{v}_n = f_n(v_n)$ are not exactly the same at each site are also feasible.

## 5  Diffusive networks with two-variable units

The aim of this section is to apply the differential inequality of the type (3.11) to networks of bistable elements described by two variables per unit. Thus we suppose that the vector $x_n = (v_n, w_n) \in \mathbf{R}^2$ and the function $f(x_n)$ introduced by eq. (2.1) has two components $f(x_n) = (h(x_n), g(x_n))$. As the two stationary states $x^{(i)}$, $(i = 0, 1)$ of eq. (2.1) are supposed to be stable, the product of the two eigenvalues of $\mathrm{D}f(x^{(i)})$ is positive, and thus:

$$A^{(i)} = \det \mathrm{D}f(x^{(i)}) > 0 \,. \tag{5.1}$$

On the other hand, we suppose here that the units of the network are coupled with the discrete diffusion operator defined by eq. (2.3) in a $d$-dimensional lattice, and we assume that the diffusion coefficients of the variables $v_n$ and $w_n$ are in the ratio $1\!:\!\delta$, respectively.

In order to fix the ideas, we rewrite eq. (2.2) with the present hypotheses, and in the case of $d = 1$, i.e., for a chain of bistable units:

$$\begin{aligned}
\frac{\mathrm{d}v_n}{\mathrm{d}t} &= h(v_n, w_n) + \alpha \left(v_{n+1} - 2v_n + v_{n-1}\right) , \\
\frac{\mathrm{d}w_n}{\mathrm{d}t} &= g(v_n, w_n) + \alpha\delta \left(w_{n+1} - 2w_n + w_{n-1}\right) ,
\end{aligned} \tag{5.2}$$

with $n \in \mathbf{Z}$.

Our analysis is valid, however, for arbitrary $d$.

**Example 5.1** We consider the FitzHugh-Nagumo model [6] which is a simple two-variable model of bistable dynamics which generalises example 4.1. In this case the functions $h$ and $g$ can be defined as

$$h(v,w) = p(v) - w \,, \quad g(v,w) = \epsilon(v - \gamma w) \,, \tag{5.3}$$

where $p(v)$ is a 'cubic-shape' function, e.g., $p(0) = p(\eta) = p(1) = 0$, $p(v) < 0$ for $0 < v < \eta$, and $p(v) > 0$ for $\eta < v < 1$. The parameter $\epsilon$ characterizes the time-scale difference in the dynamics of variables $v_n$ and $w_n$, and the parameter $\gamma$ is chosen such that the curves defined by $h(v,w) = 0$ and $g(v,w) = 0$ in the plane $(v,w)$ have three distinct intersections.

The eq. (5.2)-(5.3), with $\delta = 0$ and $\epsilon < 1$, are used in theoretical biology for modelling pulse propagation along the membranes of nerve cells [6]. Note that when $\delta = 0$, the study of the equilibrium states of (5.2) reduces to a one-variable problem, as $w_n$ can be eliminated using $g(v_n, w_n) = 0$. So to have genuinely two-variable problem we take $\delta \neq 0$. This is relevant to several interesting problems.

Now let us take a binary indexed set $(i_k)_{k \in \mathbf{S}}$, $(i_k = 0$ or $1)$, and the corresponding stationary state $X_0$ [eq. (2.4)] of the network defined by eq. (5.2) in the uncoupled limit $\alpha = 0$. By theorem 2.1, there exists a locally unique continuation $X(\alpha)$ of $X_0$ for $\alpha$ small. We will estimate a lower bound on the coupling parameter $\alpha$ for which the stationary state $X(\alpha)$ persists. We will restrict attention to $\alpha \geq 0$ as this is the case of main physical interest.

Following the procedure described in section 3, we denote the right-hand side of eq. (5.2) by the function $G(X, \alpha)$. Then we decompose $\mathrm{D}G$ according to eq. (3.3) and we denote $\mathrm{D}F$ by $\Lambda$. Here, each block of the operator $\Lambda$ is a $2 \times 2$ matrix defined as

$$\Lambda_n = \begin{pmatrix} h_v & h_w \\ g_v & g_w \end{pmatrix} \,, \quad n \in \mathbf{S} \,. \tag{5.4}$$

In this expression, the subscripts $v$ and $w$ denote partial derivatives, e.g., $h_v = \frac{\partial h}{\partial v}(x_n)$. If each block $\Lambda_n$ is invertible, so is $\Lambda$ and we can write

$$\begin{aligned} \| \Lambda^{-1} \|^{-1} &= \left( \sup_{n \in \mathbf{S}} \| \Lambda_n^{-1} \| \right)^{-1} \\ &= \min( \inf_{n \in S_0} \| \Lambda_n^{-1} \|^{-1}, \inf_{n \in S_1} \| \Lambda_n^{-1} \|^{-1} ) \end{aligned} \tag{5.5}$$

where $S_0$ and $S_1$ form the partition of $\mathbf{S}$ induced by the indexed set $\{i_k\}_{k \in \mathbf{S}}$, i.e., $k \in S_0$ if $i_k = 0$ and $k \in S_1$ if $i_k = 1$.

We are free to choose the norm in the two-dimensional state space for each unit. The estimates we will obtain on $\alpha$ will depend on the choice of norm. We choose:

$$|(v, w)| = \max(|v|, |w|),$$

and consequently the matrix norm is chosen as:

$$\| M \| = \sup_i \sum_j |M_{ij}|.$$

With this choice of norm we can write explicitly (for $i = 0$ or $1$):

$$\inf_{n \in S_i} \| \Lambda_n^{-1} \|^{-1} = \inf_{n \in S_i} \frac{|A(v_n, w_n)|}{\max(|h_w| + |g_w|, |h_v| + |g_v|)}, \tag{5.6}$$

with $A(v_n, w_n) = \det \Lambda_n$. Now, this expression must be bounded in order to achieve an inequality of the form (3.8). The idea is to use the mean value theorem for the numerator and for the denominator of eq. (5.6).

First, from eq. (5.1) it is seen that $A(v^{(i)}, w^{(i)}) = A^{(i)}$ is positive. Hence, taking into account the condition on eq. (3.7), $A(v_n, w_n)$ must remain positive over the range of $\alpha$ value considered. Therefore we can drop the absolute value of $A$ in expression (5.6). Now, $A$ can be expanded as a Taylor series in its variables such that

$$A(v_n, w_n,) = A^{(i)} + A_v^{(i)} (v_n - v^{(i)}) + A_w^{(i)} (w_n - w^{(i)}) + R^{(i)}, \tag{5.7}$$

where $R^{(i)}$ is a remainder of order 2. We shall assume that $R^{(i)}$ is positive, which must be checked for given functions $h$ and $g$. In particular, it is true for the FitzHugh-Nagumo model [eq. (5.3)]. (If $R^{(i)}$ were not positive, we would have to estimate it in terms of $\zeta$.) Then, using the variable $\zeta$ introduced by eq. (3.1) we can bound $A(v_n, w_n)$ by

$$A(v_n, w_n) \geq A^{(i)} - B^{(i)} \zeta, \tag{5.8}$$

where $B^{(i)} = |A_v^{(i)}| + |A_w^{(i)}|$.

Second, the denominator of eq. (5.6) is also expanded as a power series, and $\zeta$ is substituted such that we obtain

$$|h_v| + |g_v| \leq C_1^{(i)} + D_1^{(i)}\zeta ,$$
$$|h_w| + |g_w| \leq C_2^{(i)} + D_2^{(i)}\zeta ,$$

where $C_1^{(i)} = |h_v^{(i)}| + |g_v^{(i)}|$, $C_2^{(i)} = |h_w^{(i)}| + |g_w^{(i)}|$, and $D_1^{(i)} = |h_{vv}^{(i)}| + |g_{vv}^{(i)}| + |h_{vw}^{(i)}| + |g_{vw}^{(i)}|$, $D_2^{(i)} = |h_{vw}^{(i)}| + |g_{vw}^{(i)}| + |h_{ww}^{(i)}| + |g_{ww}^{(i)}|$. These inequalities hold if the absolute value of the partial second derivatives $|h_{vv}(x^{(i)}+\theta\Delta x)|, |h_{vw}(x^{(i)}+\theta\Delta x)|$, etc..., with $\theta \in [0,1]$ and $\Delta x = x_n - x^{(i)}$, are decreasing functions of $\theta$. Again, this property is verified in the particular case of the FitzHugh-Nagumo model. We introduce the notation

$$C^{(i)} + D^{(i)}\zeta = \max(C_1^{(i)} + D_1^{(i)}\zeta, C_2^{(i)} + D_2^{(i)}\zeta) . \tag{5.9}$$

Finally, the substitution of eq. (5.8)-(5.9) into eq. (5.5)-(5.6) leads to

$$\|\Lambda^{-1}\|^{-1} \geq \min_{i=0,1} \frac{A^{(i)} - B^{(i)}\zeta}{C^{(i)} + D^{(i)}\zeta} . \tag{5.10}$$

We could insert this estimate into (3.7), but it would lead to a more complicated form of equation than (3.11). So instead we will bound (5.10) by

$$\|\Lambda^{-1}\|^{-1} \geq \min_{i=0,1}(a_0^{(i)} - a_1^{(i)}\zeta) , \tag{5.11}$$

where $a_0^{(i)} = A^{(i)}/C^{(i)}, a_1^{(i)} = (A^{(i)}D^{(i)} + B^{(i)}C^{(i)})/C^{(i)2}$, and $b_1 = (PC^{(i)} - A^{(i)}Q)/C^{(i)2}$. The reader can check this by multiplying both expressions by $C^{(i)} + D^{(i)}$.

At this stage, we have obtained the inequality (3.8). In order to achieve the desired inequality (3.11) we have yet to determine the factors $b_0, c_0$ and $c_1$ defined by eq. (3.9)-(3.10). In the case of discrete diffusion, these factors are simply given by the following inequalities.

$$\begin{aligned}|K(X)| &= \sup_{n \in \mathbf{S}} |\sum_{k=1}^{2d} D(x_{i_k}(\alpha) - x_n(\alpha))| \\ &\leq 2d \, \|D\| \sup_{m,n \in \mathbf{S}} |x_m(\alpha) - x_n(\alpha)| \\ &\leq 2d \, \|D\| \, (\xi + 2\zeta) ,\end{aligned} \tag{5.12}$$

where $\xi$ is equal to $|x^{(1)} - x^{(0)}|$.

Finally, for the discrete diffusion operator (2.3) it can readily be verified that the norm of $DK$ is given by

$$\|DK\| = 4d \, \|D\| . \tag{5.13}$$

In conclusion, for a diffusive network of bistable elements described by two variables per unit we obtain the differential inequality

$$\left|\frac{\mathrm{d}\zeta}{\mathrm{d}\alpha}\right| \leq \frac{2d\,\|D\|\,(\xi + 2\zeta)}{a_0 - a_1\zeta - 4d\,\|D\|\,|\alpha|}\,. \tag{5.14}$$

which is of the form (3.11). Thus we can use (3.12) and obtain

$$\alpha_0 = \frac{a_0^2}{4d\,\|D\|\,a_1\xi} + \mathcal{O}(\frac{a_0^3}{a_1^2\xi^2}) \tag{5.15}$$

*The FitzHugh-Nagumo system*

Let us deduce from (5.15) a lower bound of the critical coupling for which steady states persist in the network for the FitzHugh-Nagumo model defined by eq. (5.2)-(5.3). It is convenient to divide the second equation of the system (5.2) by $\delta$, and this leads to better estimates. In this case the various factors of the inequality (5.10) are easily calculated and we can write

$$\|\Lambda^{-1}\|^{-1} \geq \min_{i=0,1}\left(\frac{-\epsilon\gamma p_v^{(i)} - \epsilon\gamma|p_{vv}^{(i)}|\zeta}{\max(\delta + \epsilon\gamma, |p_v^{(i)}|\delta + |p_{vv}^{(i)}|\zeta\delta + \epsilon)}\right)\,. \tag{5.16}$$

Let us treat the case where the first term in the denominator is the largest one (which will be the case for all $\zeta$ of interest if $\gamma$ is large enough). Following eq. (3.8) we bound (5.16) by

$$\|\Lambda^{-1}\|^{-1} \geq a_0 - a_1\zeta\,, \tag{5.17}$$

where $a_0 - a_1\zeta$ is defined by $\min_{i=0,1}(-p_v^{(i)} - |p_{vv}^{(i)}|\zeta)\epsilon\gamma/(\delta + \epsilon\gamma)$.

Now, $\|D\| = 1$. The remaining constants to insert in (5.15) are $a_0, a_1$ and $\xi$. These depend on the form of the function $p(v)$. As a concrete application, we consider the case where $p(v)$ is a cubic polynomial defined by eq. (4.10). Then the equilibrium states of the bistable unit are at $(0,0)$ and $(\xi, \xi/\gamma)$ with

$$\xi = \frac{1}{2}\left(1 + \eta + \sqrt{(1-\eta)^2 - \frac{4}{\gamma}}\right) \simeq 1 - \frac{1}{(1-\eta)\gamma}\,.$$

Furthermore, in the present case we have $p_v^{(0)} = -\eta, p_v^{(1)} = \eta - 1, p_{vv}^{(0)} = 2(\eta+1)$ and $p_{vv}^{(1)} = -2(2-\eta)$. So, if $\eta < \frac{1}{2}$, it can be shown that eq. (5.17) can written with

$$a_0 = \frac{\epsilon\gamma\eta}{\delta + \epsilon\gamma}\,, \qquad a_1 = \frac{2(\eta+1)\epsilon\gamma}{\delta + \epsilon\gamma}\,,$$

Introducing the parameters $a_0, a_1$ in eq. (5.15) we obtain that for the FitzHugh-Nagumo model continuation from the uncoupled limit is possible for $\alpha < \alpha_0$ with

$$\alpha_0 = \frac{\epsilon\gamma\eta^2}{8d(\delta + \epsilon\gamma)(\eta + 1)\xi} + \mathcal{O}(\frac{\eta^3}{(1+\eta)^2\xi^2}) \;.$$

We note that in the case $\gamma \to \infty$ one recovers, at lowest order, the value $\alpha_0 \simeq \eta^2/8$ which we obtained in eq. (4.11) for the one-variable cubic Nagumo model.

## 6  Optical bistability

In this section we will apply the scheme proposed in section 3 to an example taken from nonlinear optics. As mentioned in the introduction, one can conceive optical devices which, when submitted to an external driving field, may exhibit two different steady states. The first stable stationary state is characterized by a low transmission rate whereas the second stationary state has a high transmission rate. Thus, endowed with its two states, each device can cope with a unit of information, i.e., a *bit*. Therefore, a simple idea is to consider a network of such optical devices in order to create an optical memory. However, one of the problems which could arise in this context is that when the optical devices are brought close to each other, they become coupled, at least by means of their overlapping evanescent fields. The coupling might prevent the network from having stable stationary patterns. In this context, we propose in the present section to show how a bound on the coupling strength can be estimated, which guarantees the persistence of the stationary patterns which would exist in the absence of coupling. We note that this issue has been addressed by Firth in ref.[9], using a method closely related to the one of Keener [8].

As a simple model, we consider a network of identical optical ring cavities containing a nonlinear absorptive medium, and coupled in a $d$-dimensional lattice by means of their field amplitude. We assume that the dynamics of the network is governed by the following equations:

$$\begin{aligned}
\dot{v}_n &= -v_n + y - 2Cw_n + \alpha\left(v_{n+1} - 2v_n + v_{n-1}\right), \\
\dot{w}_n &= \frac{\gamma_\perp}{\kappa}(-w_n + v_n z_n), \\
\dot{z}_n &= \frac{\gamma_\parallel}{\kappa}(-z_n + 1 - v_n w_n),
\end{aligned} \quad (6.1)$$

with $n \in \mathbf{Z}$. For simplicity the equations have been written with $d = 1$.

In the absence of coupling ($\alpha = 0$), the eq. (6.1) represent a simple three-variable model for studying absorptive optical bistability [14]. This model can be derived from the Maxwell-Bloch equations in a semi-classical mean-field approximation applied to a set of broadened two-level atoms [14]. The variables $v_n, w_n$, and $z_n$ are reduced variables corresponding respectively to the real field amplitude produced in the cavity, the atomic polarization and the population difference between the lower and upper atomic levels. The model is also characterized by five parameters amongst which $\kappa, \gamma_\perp$ and $\gamma_\parallel$ are various decay rates. The amplitude of the field injected into the ring-cavity is represented by $y$ and $C$ corresponds to a combination of several physical parameters and is called the bistability parameter. For more details, see e.g. ref [14].

Although the bistable units are described by three variables, when these are coupled only through the variable $v$, the equilibrium states of the network can be studied by solving a one-variable system of the form (4.1) with

$$f(v_n) = -v_n + y - 2C \frac{v_n}{1 + v_n^2} \ .$$

Indeed, in the case where only $v$ is coupled, the stationary states of $w_n$ and $z_n$ can be expressed as function of $v_n$, namely $w = v(1+v^2)^{-1}$ and $z = (1+v^2)^{-1}$. Therefore the scheme of section 4 apply. First we analyse the roots of $f(v)$. In order to deal with analytical expressions for the stationary states we choose the value of parameter $y$ as

$$y = C + 1 \ .$$

Then the stable equilibrium states of $f$ are determined as

$$v^{(0)} = 1 \ , \tag{6.2}$$

$$v^{(1)} = \frac{1}{2} \left( C + \sqrt{C^2 - 4C - 4} \right) \simeq C - 1 \ . \tag{6.3}$$

Second we analyse the critical points of function $f$, i.e., the roots of $f'(v)$. These are

$$c_\pm = \sqrt{C - 1 \pm \sqrt{C(C-4)}} \ .$$

Now, as suggested in ref. [14], it is interesting to consider the limit of "fully developed hysteresis cycle", that is the limit of large values of $C$. In this case the nearest critical points to $\{v^{(0)}, v^{(1)}\}$ is $c_-$ because $|c_- - v^{(0)}| = \mathcal{O}(C^{-1})$ whereas $|c_+ - v^{(1)}| = \mathcal{O}(C)$. However, as $|f'(v^{(0)})| > |f'(v^{(1)})|$, we are in a situation where there is an interval $[0, \zeta_m]$ about which $|f'(v^{(0)} + \zeta_m)| > |f'(v^{(1)} - \zeta_m)|$. This case is analogous to the one described by eq. (4.8). Hence following eq. (4.9) –or a slight generalisation when $(v^{(0)}, v^{(1)})$ is not normalized to $(0, 1)$– continuation of the stationary states from the uncoupled limit is possible for $\alpha < \alpha_0$ where

$$\alpha_0 = \frac{1}{2(v^{(1)} - v^{(0)})}(-f(c_-) + f(v^{(0)} + \zeta_m) + f(v^{(1)} - \zeta_m)). \tag{6.4}$$

The analytical expression of $f(c_-)$ is a complicated function of $C$ we can expand as a power series in $C^{-1}$ and we get

$$f(c_-) = -\frac{1}{C} + \mathcal{O}(C^{-2}).$$

On the other hand, the terms in (6.4) including $\zeta_m$ can be shown to be of order $C^{-3}$. Hence, at the lowest order in $C^{-1}$, eq. (6.4) gives

$$\alpha_0 = \frac{1}{2C^2} + \mathcal{O}(C^{-3}).$$

This value of the coupling strength between the optical bistable cavities provides a lower bound up to which the stationary patterns (2.4) continue to exist. Therefore this value of $\alpha_0$ gives a bound for which the optical network has a family of stationary states representing arbitrary indexed sets of binary digits. This is a prerequisite for designing optical memories.

Finally note that we could consider diffusion of more than one variable in system (6.1). In this multivariable problem we could apply method as in section 3-5.

## 7 Stability of the steady states

In this section we prove the perhaps obvious result that the continuation of a stable steady state of the uncoupled network is stable for small $\alpha$. A word of caution is in order, however: it is not necessarily true that the stability-type is preserved for all $\alpha$ to which a steady state can be continued. It is perfectly possible that the equilibrium might undergo a Hopf bifurcation at some $\alpha$ : the equilibrium persists but changes stability-type.

We treat spectral stability only. For a steady state $X(\alpha)$, we write $M(\alpha) = DG(X(\alpha), \alpha)$. The idea is that since $M(\alpha)$ varies continuously with $\alpha$, its spectrum can not move too fast. To make it precise, we introduce the resolvent of $M$:

$$R_\lambda(M) = (\lambda I - M)^{-1} ,$$

defined for $\lambda \in \mathbf{C} \setminus \operatorname{spec} M$. Then, using the formula $A^{-1} - B^{-1} = A^{-1}(B - A)B^{-1}$, one proves that $\lambda \notin \operatorname{spec} M'$ for all $M'$ such that

$$\| M' - M \| < \| R_\lambda(M) \|^{-1} . \tag{7.1}$$

Now, let

$$r = \sup_{\operatorname{Re} \lambda \geq 0} \| R_\lambda(M(0)) \| .$$

This is easy to calculate, as $R_\lambda(M(0))$ is block-diagonal, and by our hypotheses on $\operatorname{spec} DF(x^{(i)})$ and boundedness of $M$, $r < \infty$. Then it follows from (7.1) that $M(\alpha)$ has no spectrum in the right half plane (nor on the imaginary axis) as long as

$$\| M(\alpha) - M(0) \| < \frac{1}{r} .$$

Note that we can also continue unstable steady states of the uncoupled network for small $\alpha$. They remain unstable for small enough $\alpha$, because by the above, the resolvent remain bounded on the semicircle $\Gamma$ with diameter $(-iB, +iB)$ on the imaginary axis, lying in the closed right half plane, for small $\alpha$, where $B$ is any number such that

$$\| M(0) \| < B .$$

Then there must continue to be spectrum inside the semicircle $\Gamma$ because the spectral projection operator $P_\Gamma = \frac{1}{2\pi i} \int_\Gamma R_\lambda(M(\alpha)) d\lambda$ depends continuously on $\alpha$ and is non-zero for $\alpha = 0$.

## 8 Conclusion

The aim of this paper was to study the stationary states of networks of bistable units with weak coupling. We proved that, for a large class of network interaction, not necessarily linear, and for small values of the coupling strength, the network presents a high steady state multistability inherited from the bistable behaviour of each unit. In some cases, it is possible to give a bound on the coupling strength up to which the stationary solutions continue to exist. We have applied this analysis to concrete examples, such as the two-variable FitzHugh-Nagumo system and a simple model describing an instance of optical bistability.

It is plausible that the existence of so many stable stationary states would block wavefronts from propagating in the network. This phenomenon is called 'propagation failure', and has been shown to be an important discreteness effect in diffusion-coupled bistable systems [8]-[12]. We wish to emphasize that our present results, however, do not say anything about the existence or not of propagating waves or fronts. Nevertheless, we believe that our work could be extended to prove structural stability of the uncoupled case, i.e. topological equivalence of the weakly coupled case to the uncoupled case. This would preclude existence of propagating waves of interest, as the uncoupled case has no propagating wave solutions, except those involving the unstable steady state between the two stable steady states. These are probably not of any interest as they require preparation of the initial condition in a multiply-unstable situation.

The present study illustrates a simple and efficient method for investigating weakly coupled systems. The method consists in analysing, first of all, the solutions of the system in the limit of vanishing coupling in the network. This limit is interesting because it may provide an infinite number of explicit solutions each of which having possibly a unique continuation for non-zero value of the coupling. Moreover, it is possible, in principle, to estimate the range of coupling strength over which the continuation from the uncoupled limit exists.

Note that, as in [7,13], we could also deduce a localisation length for the steady states for small $\alpha$ though we have not done so here. This means that the effect of changing the state of one unit decays exponentially from that site. For small $\alpha$, and local coupling, the decay rate is $\mathcal{O}(\alpha)$.

The procedure proposed in this study could probably be applied to more complex situations than stationary states. For example, the units of the network could be replaced by bistable oscillators which have a stable stationary state in addition to a stable limit-cycle. In that case it is interesting to determine whether there exist stable localized oscillations, as a continuation from the uncoupled limit. In fact this question has already been addressed and answered affirmatively in the case of Hamiltonian or time-reversible oscillatory networks [13]. For such systems, it has been recently proved that there exist 'discrete breathers', i.e., localized oscillatory solutions in the network, when the coupling strength between the oscillators is sufficiently small. Notice that in that case, the operator $G$ defined by eq. (2.5) corresponds to an operator de-fined on 'spaces of $T$-periodic loops', in order to cope with the time-dependent behaviour of the oscillators. For dissipative networks, the sort of results we could expect would be a network analogue of the theory of persistence of normally hyperbolic invariant manifolds.

**Acknowledgement**

This work was supported by an E.C. Human Capital and Mobility Network grant on 'Nonlinear Phenomena and Complex Systems', and an ECHCM net on 'Nonlinear approach to coherent and fluctuating phenomena in condensed matter and optical physics'.